\documentclass{aa}
\usepackage{txfonts}
\usepackage{amsmath}
\usepackage{color}
\usepackage{xcolor}
\usepackage[normalem]{ulem}
\usepackage{multicol}
\usepackage[colorlinks=true,citecolor=blue,linkcolor=blue]{hyperref}
\usepackage{textcomp}
\usepackage{subcaption}
\usepackage{graphicx}
\usepackage{float}
\usepackage{placeins}
\usepackage{multirow}
\usepackage{changepage}
\usepackage{colortbl}
\usepackage{amsfonts}
\usepackage{booktabs}
\usepackage{siunitx}
\usepackage{amssymb} 
\usepackage{enumitem}
\usepackage{multirow}

\author{
J.M.G.H.J. de Jong\inst{\ref{inst:Leiden}} 
\and R.J. van Weeren\inst{\ref{inst:Leiden}}
\and T.J. Dijkema\inst{\ref{inst:Astron}}
\and J.B.R. Oonk\inst{\ref{inst:surf},\ref{inst:Leiden}}
\and H.J.A. R\"ottgering\inst{\ref{inst:Leiden}}
\and F. Sweijen\inst{\ref{inst:Durham}}
}

\institute{
Leiden Observatory, Leiden University, PO Box 9513, 2300 RA Leiden, The Netherlands\relax\label{inst:Leiden} 
\and 
SURF/SURFsara, Science Park 140, 1098 XG Amsterdam, The Netherlands\relax\label{inst:surf} 
\and 
ASTRON, The Netherlands Institute for Radio Astronomy, Oude Hoogeveensedijk 4, 7991 PD Dwingeloo, The Netherlands\relax\label{inst:Astron}
\and
Centre for Extragalactic Astronomy, Department of Physics, Durham University, Durham, DH1 3LE, UK\relax\label{inst:Durham}
}

\begin{document}

\title{Unlocking ultra-deep wide-field imaging with sidereal visibility averaging}

\date{Received XXX 2024 / Accepted XXX 2024}

\begin{abstract} 
    {Producing ultra-deep high-angular-resolution images with current and next-generation radio interferometers introduces significant computational challenges. In particular, the imaging is so demanding that processing large datasets, accumulated over hundreds of hours on the same pointing, is likely infeasible in the current data reduction schemes. In this paper, we revisit a solution to this problem that was considered in the past but is not being used in modern software: sidereal visibility averaging (SVA). This technique combines individual observations taken at different sidereal days into one much smaller dataset by averaging visibilities at similar baseline coordinates. We present our method and validated it using four separate 8-hour observations of the ELAIS-N1 deep field, taken with the International LOw Frequency ARray (LOFAR) Telescope (ILT) at 140~MHz. Additionally, we assessed the accuracy constraints imposed by Earth's orbital motion relative to the observed pointing when combining multiple datasets. We find, with four observations, data volume reductions of a factor of 1.8 and computational time improvements of a factor of 1.6 compared to standard imaging. These factors will increase when more observations are combined with SVA. For instance, with 3000~hours of LOFAR data aimed at achieving sensitivities of the order of $\mu$Jy beam$^{-1}$ at sub-arcsecond resolutions, we estimate data volume reductions of up to a factor of 169 and a 14-fold decrease in computing time using our current algorithm. This advancement for imaging large deep interferometric datasets will benefit current generation instruments, such as LOFAR, and upcoming instruments such as the Square Kilometre Array (SKA), provided the calibrated visibility data of the individual observations are retained.}
    \maketitle
\end{abstract}

\section{Introduction}\label{sec:introduction}

Modern and upcoming radio interferometers such as the International LOw Frequency ARray (LOFAR) Telescope \citep[ILT][]{haarlem2013} and the Square Kilometre Array \citep[SKA][]{dewdney2009} and its pathfinders \citep{schinckel2012, tingay2013, jonas2016} advance our ability to study the universe in unprecedented detail.
However, the substantial volumes of data generated by these instruments present significant challenges for data processing and storage (costs). This issue is especially pronounced when combining multiple observations of the same sky area for deep high-resolution imaging, as this involves processing much larger volumes of data compared to imaging with a single observation or at lower resolutions. To ensure that the high costs associated with data processing do not outweigh the scientific benefits, it is crucial to employ more efficient data processing and handling techniques.

Recent deep surveys dedicated to specific areas on the sky, each spanning hundreds of hours with LOFAR, have produced wide-field images with sensitivities reaching 20~$\mu$Jy~beam$^{-1}$ at 144MHz and 6\arcsec~resolution \citep{kondapally2021, duncan2021, tasse2021, sabater2021, best2023, bondi2024}. The calibration and imaging were conducted using only the Dutch core and remote stations, excluding all international stations. Including also the international LOFAR stations allows for the creation of ultra-deep, degree-scale images with sub-arcsecond resolutions, due to the additional baselines extending up to 2000~km. These images contain, at sub-arcsecond~resolutions, up to 10 billion pixels over 2.5$\times$2.5 degrees and are generated from data amounting to tens to hundreds of terabytes when combining all available observations of the same field. The feasibility of generating these degree-scale images after calibrating all LOFAR's international stations has been proven by producing the first 0.3\arcsec~wide-field image with a sensitivity of 32~$\mu$Jy~beam$^{-1}$ \citep{sweijen2022}. Recent advancements have led to the production of the deepest wide-field image currently available at 0.3\arcsec~resolution and 140\,MHz, with a sensitivity of 14~$\mu$Jy~beam$^{-1}$, by combining four observations \citep{dejong2024}. While these pioneering studies have progressively developed strategies that can be scaled up to process hundreds of observing hours of the same pointing, they have also highlighted that the final imaging of the calibrated data is a computational bottleneck. This step consumes about 80\% of the total computational time of the entire data reduction pipeline, due to the large data volume left after calibration. To address this, a reduction of the data volumes before imaging without significantly compromising image quality is necessary.

An effective way to compress the data volume of interferometric data is by decreasing the number of visibilities that are needed to create an image. These visibilities are measurements of the correlated signals between pairs of antennas (baselines), capturing the Fourier components of the sky brightness distribution essential for image reconstruction. One commonly used method to do this is by using baseline-dependent averaging \citep[BDA; e.g.][]{cotton1986, cotton2009, skipper2014, wijnholds2018, atemkeng2022}. This leverages the fact that different baseline lengths may have different time and frequency resolutions to recover the information for imaging without introducing time and bandwidth smearing. For short baselines, which measure large-scale structures in the sky, the visibility function changes slowly with time and frequency. In contrast, long baselines are sensitive to small spatial resolutions, where the visibility function may vary more rapidly with time and frequency. Therefore, visibilities for short baselines can be averaged over longer time periods and broader frequency channels compared to longer baselines. BDA can, for example, reduce data volumes for the SKA-Low and SKA-Mid up to about \textasciitilde85\% \citep{wijnholds2018, deng2022}. Alternatively, it is also possible to perform data volume compression by using Dysco compression \cite{offringa2016}. This method does not reduce the number of visibilities but uses lossy compression to reduce the storage space and, thus, the reading speed of visibility data. Dysco compression works well on lower signal-to-noise ratio (S/N) data by taking advantage of the fact that variations in the data are primarily due to Gaussian noise rather than the actual signal. This allows for efficient data compression without losing important visibility information to reconstruct images. On average, this technique reduces the data volume for a typical LOFAR observation by a factor of 4 to 6, partly also due to more efficient storage of weights. 

Despite the successes in terms of compressing datasets of individual observations, data volume and computational time still scale linearly when performing deep imaging with multiple observations, due to the increased number of visibilities that in imaging software all need to be processed.
One way to reduce computational wall time is to create separate images from individual observations and then average them in image space. While this approach enables parallel processing for each observation, it still necessitates gridding all visibilities from the different observations during imaging, resulting in about the same total computational time as imaging all visibilities together (which we refer to as `standard imaging' throughout this paper). Additionally, image quality is worse when averaging images compared to standard imaging because deconvolution of faint sources can only be done when performed with all observations together.
An alternative approach to combine observations and reduce data volume and computational time is to utilize the repeating baseline tracks from deep observations taken over multiple sidereal days. In this way, it is possible to reduce the number of visibilities that need to be imaged. This was, for instance, conducted by \cite{owen2008}, who averaged visibilities at the same hour angles using data from the Very Large Array (VLA) at 1.4 GHz with a total integration time of 140~hours. They implemented this method in the Astronomical Image Processing System \citep[\texttt{AIPS}][]{vandiepen1994, glendenning1996, greisen2003} as the task \texttt{STUFFR}\footnote{\url{http://www.aips.nrao.edu/cgi-bin/ZXHLP2.PL?STUFFR}}, which combines several \texttt{AIPS} tasks to perform visibility averaging over different sidereal days. This was later also used for deep imaging of the GOODS-N field with the VLA \citep{owen2018}.


\texttt{AIPS} was originally designed to handle smaller data volumes compared to the large datasets produced by modern interferometers, such as LOFAR and the SKA. While some processing improvements have been developed for \texttt{AIPS} \citep[e.g.][]{kettenis2006, cotton2008, bourke2014}, it is not optimized to deal with direction-dependent effects (DDEs). These pose an added challenge to achieving wide-field images with high dynamic range and S/N below a few hundred MHz. DDEs are primarily introduced by the ionosphere and beam model errors, distorting the `real' visibilities differently across the field of view. For LOFAR and SKA data processing have been developed, capable of efficiently processing datasets of the order of tera-to petabytes on large powerful multi-CPU machines, while DDEs may be corrected with software packages such as \texttt{DP3}\footnote{\url{https://dp3.readthedocs.io}} \citep{dp3, dijkema2023}, \texttt{SPAM} \citep{intema2009}, \texttt{Sagecal}\footnote{\url{https://github.com/nlesc-dirac/sagecal}} \citep{sage2011}, \texttt{KillMS}\footnote{\url{https://github.com/saopicc/killMS}} \citep{tasse2014, tasse2014b, smirnov2015}, and \texttt{facetselfcal}\footnote{\url{https://github.com/rvweeren/lofar_facet_selfcal}} \citep{vanweeren2021}. These are typically integrated in a facet-based approach, where the field is divided into multiple facets \citep{vanweeren2016, williams2016}, with each facet receiving its own calibration solutions. Since calibration for systematic and ionospheric effects is best performed on a per-observation basis, it is essential to average visibilities from different observations corresponding to different sidereal days only after calibrating for the DDEs in each facet. This requires, before averaging visibilities from different observations, to split off datasets for each facet from the full dataset and treating each facet separately, as was demonstrated by \cite{sweijen2022} and \cite{dejong2024} without averaging visibilities for similar hour angles or baseline coordinates.

In this paper, we are exploring a revised method to average visibilities over sidereal days on already calibrated facet data. We term this `sidereal visibility averaging' (SVA). To demonstrate our method, we average visibilities from datasets from four different LOFAR observations calibrated by \cite{dejong2024}. By testing various averaging settings, we compare and optimize the balance between image quality and computing costs. We also consider effects on the binning of similar baseline coordinates and frequency offsets due to Earth's celestial motion. Given that imaging accounts for approximately 75-80\% in the current LOFAR sub-arcsecond wide-field data reduction pipeline \citep{sweijen2022, dejong2024}, we address a significant part of current computational challenges for ultra-deep imaging of a single pointing on the sky. While we focus on data from LOFAR in this paper, we advocate this method as a viable solution to address computing and storage challenges for deep multi-epoch imaging with other instruments as well.

In Section \ref{sec:algorithm_section}, we first discuss the SVA algorithm. This is followed by an overview of the data used in this paper in Section \ref{sec:data}. We then present our results in Section \ref{sec:results}, followed by a discussion in Section \ref{sec:discussion}. Finally, we conclude with a summary and conclusions in Section \ref{sec:conclusion}.

\section{Sidereal visibility averaging}\label{sec:algorithm_section}

Interferometric datasets consist of several components, including a time axis, a frequency axis, baseline coordinates (\textit{uvw}), visibilities, and their corresponding weights. The initial time and frequency resolution of these axes are determined by the settings of the interferometer or correlator. Visibilities are the measurements of the correlated signals for baselines at specific moments in time and for different frequencies. As the Earth rotates, these baseline tracks move over time, with each timestamp corresponding to a point in the \textit{uvw} plane \citep[e.g.][]{brouw1975}. This plane is a coordinate system used to describe the relative positions of antennas. When observations with the same pointing centre are conducted over more than one sidereal day, parts of the baseline tracks are repeated. This repetition allows us to average the visibilities in similar baseline coordinate bins. We discuss in this section how the SVA algorithm makes efficiently use of the possibility to average observations over multiple sidereal days. In this paper, we use LOFAR as an example because we have recently reduced high-resolution LOFAR data at hand \citep{dejong2024}.

\subsection{Frequency and time axis}\label{sec:freqsid_axis}

The detected radio signals are split into specific frequency bands using a polyphase filterbank, which processes and organizes these signals into channels. For LOFAR, the central frequency of these channels is pre-defined which ensures that LOFAR data is stored using the same channel centres in the LOFAR Long Term Archive (LTA)\footnote{\url{lta.lofar.eu}} \citep{haarlem2013}. As frequencies may be averaged during processing, it is essential that these channels are averaged by common denominators, such that when applying SVA, we only have to match the corresponding frequency channels of the visibilities. Matching visibilities with frequency offsets can otherwise result in image distortions. This may include inaccurate source positioning when channels with different centres are combined or bandwidth smearing if observations with different or too large frequency channel widths are combined. However, in the context of SVA -- where adjustments to the \textit{uvw}-plane are only influenced by the time axis -- frequency-related effects on image accuracies do not arise, as long as SVA is applied to data with frequency channels that share the same centres and widths. It is important to note that frequency-related issues can still be introduced by Doppler shifts, as we subsequently discuss in Section \ref{sec:doppler}.

The time axis from observations taken at different moments must be converted to a common time axis, which can be done using the local sidereal time (LST). The LST relates to the hour angle relative to the vernal equinox, allowing us to align the time axes from different observations to a single sidereal day. We evaluate the LST at the canonical centre of LOFAR. The time tracks of different observations in LST do not need to overlap exactly, as observations at the same pointing are not taken simultaneously on a sidereal day. For instance, if the integration time for an observation is 2~seconds, another observation of the same pointing could be offset by 1 second in sidereal time. Since the \textit{uvw}-plane and the binning in this plane between different input datasets depends on the time resolution, careful consideration of the time axis is essential in reconstructing the correct image. A critical factor is time smearing. To achieve high-quality images while minimizing data volume, it is crucial to balance time resolution with data volume and image quality. This balance is further discussed in Sections \ref{sec:imagequality_res}, \ref{sec:computing}, and \ref{sec:imagequality_dis}. Note that mapping different observations to a common LST axis leads to the loss of intrinsic time information for certain astronomical objects. This is most notable for the varying flux of transient sources.

While one day in LST consists of about 23.93~hours, we use the fact that the Fourier transform of a real-valued function (the sky brightness distribution) is Hermitian, meaning that the transform exhibits complex conjugate symmetry. Hence, a visibility at coordinates $(u,v,w)$ has a corresponding complex conjugate at $(-u,-v,-w)$. Therefore, our output dataset does not cover more than 11.97~hours.

\subsection{Algorithm}\label{sec:algorithm}

\begin{figure*}[htbp]
 \centering
\includegraphics[width=0.95\linewidth]{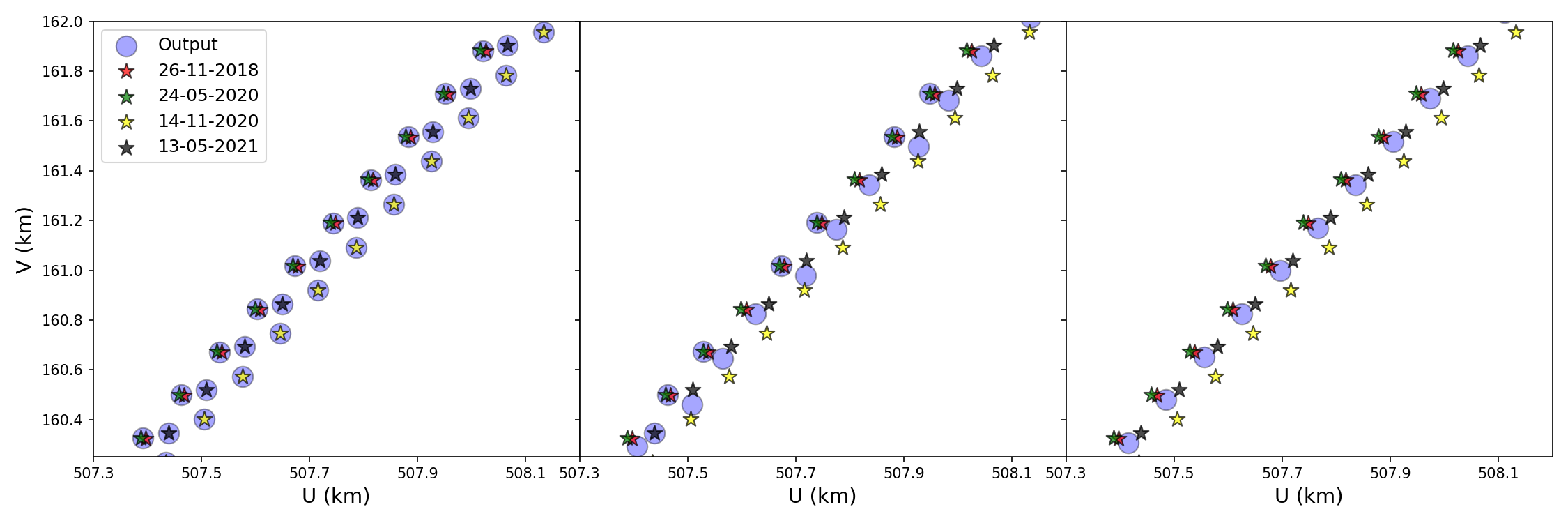}
  \caption{Different \textit{uv} samplings for one baseline with length 594\,km between a Dutch core station (CS001HBA) and the Swedish LOFAR station at Onsala Space Observatory (SE607HBA). The input datasets are represented by red, green, yellow, and black stars and labelled by their observing dates. The output dataset is represented by blue dots. The input data has a time resolution of 8 seconds. \textit{Left panel:} Output time sampling at $\Delta t=1$ second. \textit{Centre panel:} Output time sampling according to Equation \ref{eq:bridle} at $\Delta t =3.6$ seconds. \textit{Right panel:} Output time sampling at $\Delta t=8$ seconds.}
\label{fig:baseline_stack}
\end{figure*}

The SVA algorithm involves the following steps:

\begin{enumerate}
\item First, using the LST, we construct a time axis for the output dataset that encompasses all LST points from the input datasets. Though the measurement set allows specifying a time axis in LST, we convert the times to a `representative' UTC time around the median time of all observations. The time resolution ($\Delta t$) can be specified as input or can be calculated using the angular resolution ($\theta_{\text{res}}$) and maximum distance from the phase centre ($\theta$) with the following formula:
\begin{equation}\label{eq:bridle}
\Delta t = 2.9\times 10^{4}\left(\frac{\theta_{\text{res}}\sqrt{1-\tau}}{\theta}\right),
\end{equation}
where $\tau$ is the time smearing, which is equivalent to the peak intensity loss of a source. This formula is based on the time-smearing equations from \cite{bridle1999} for the average smearing effect on an image. We set $\tau$ by default equal to 0.95, but it can be changed to a more conservative value closer to 1. We also create a frequency axis that includes all frequency channels from the input datasets. 
\item Next, we add all unique LOFAR stations from the input datasets to the output dataset and make mappings that map the LOFAR station IDs from the output to those in the input datasets. This allows us to quickly identify which data entries correspond to which baseline across all the input and output datasets, as not all LOFAR observations are observed using the same set of stations.
\item Using the baseline coordinates from the input datasets and the LST from each observation, we obtain a preliminary estimate of the baseline coordinates for the output dataset by applying nearest neighbour interpolation for each baseline, using \texttt{scipy}'s \texttt{interpolate} library \citep{scipy2020}. Nearest neighbour interpolation yields similar results to other interpolation methods, as we further refine the accuracy of the output baseline coordinates by averaging all nearest \textit{uvw} values from the input datasets. Note that each baseline coordinate from an input dataset can only correspond to one baseline coordinate in the output dataset. In Figure \ref{fig:baseline_stack} we demonstrate this procedure by showing the baseline coordinates from the input dataset compared to the output dataset for one baseline and different time resolutions.
\item We create index mappings between the baseline coordinates of the input and the nearest baseline coordinates from the output datasets. These mappings allow us to quickly determine for each baseline coordinate in the input datasets to which baseline coordinate from the output dataset these correspond during averaging of the visibilities and summing the weights as explained below.
\item Visibility weights are factors that account for the reliability of visibilities, improving overall accuracy and the S/N during imaging. To obtain the visibility weights of the output dataset ($W$), we sum the visibility weights from the input datasets ($W_{i}$):
\[
\bar{W}(u,v,w) = \sum_{i} W_i(u,v,w),
\]
This ensures that the most reliable visibilities in the output dataset have the largest weights. Also, \textit{uvw} for which a subset of the input observations contribute points, has correspondingly lower weights.
\item We utilize the visibility weights to compute a weighted average of the visibility data, yielding the output visibility values as follows:
\[
\bar{V}(u,v,w) = \frac{\sum_{i} V_i(u,v,w) W_i(u,v,w)}{\bar{W}(u,v,w)},
\]
\item Finally, we add a flagging column, flagging all values with output visibility weights equal to 0, as these correspond to \textit{uvw} coordinates in the output dataset that did not have neighbouring visibilities from the input data or were already flagged in the input data.
\end{enumerate}

Since the SVA algorithm considers mappings between the same baselines of the input and output data, our method is compatible with datasets where BDA has been applied. We also ensured that the data could be compressed using Dysco compression by testing with the default settings of 10 bits per float for visibilities and 12 bits per float for visibility weights in \texttt{DP3}. We measured the image RMS noise with and without Dysco compression and found a difference of less than 0.1\%. However, it is important to note that we only tested this with four datasets (see Section \ref{sec:data}). Using more datasets or the presence of very bright sources could increase the S/N, potentially leading to image quality losses with Dysco compression. In such scenarios, it may be necessary to consider more conservative bit-rates or forgoing Dysco compression entirely.

The SVA code is currently written with Python and may be implemented as a separate step in data reduction pipelines.\footnote{The code is currently available at \url{https://github.com/jurjen93/sidereal_visibility_avg}} The code builds mainly on the functionalities from \texttt{casacore}\footnote{\url{https://casacore.github.io/python-casacore/}} \citep{casacoreteam, casateam2022} to work with measurement sets, which are the standard data format used for radio interferometric data\footnote{\url{https://casacore.github.io/casacore-notes/229.pdf}}. We also use \texttt{astropy}\footnote{\url{https://www.astropy.org}} \citep{astropy1, astropy2, astropy3} for unit conversions and the \textit{uvw} coordinate system and \texttt{scipy}\footnote{\url{https://scipy.org}} \citep{scipy2020} for nearest neighbouring interpolation and retrieving the binary-tree quick nearest neighbour lookup from \cite{maneewongvatana2002}. To increase the processing speed we also utilize \texttt{joblib} \citep{joblib}\footnote{\url{https://joblib.readthedocs.io}} for parallel processing. To increase efficiency of the algorithm, it may be investigated in the future to port the code to a more efficient programming language or implement it in an already existing efficient radio astronomical software package.

\section{Data}\label{sec:data}

\begin{table*}[htbp]
\caption{Metadata from the four used ELAIS-N1 observations.}
\centering
\begin{tabular}{lllll} \toprule
    \textbf{Observation ID} & \textbf{\textit{L686962}} & \textbf{\textit{L769393}} & \textbf{\textit{L798074}} & \textbf{\textit{L816272}} \\ \midrule
   \textbf{Observation date} & 26-11-2018 & 24-05-2020 & 14-11-2020 & 13-05-2021 \\
   \textbf{Start time} & 07:13:43 & 19:20:26 & 08:11:00 & 19:41:00 \\
   \textbf{Pointing centres} & 16:11:00, +54.57.00 & 16:11:00, +54.57.00 & 16:11:00 +55.00.00 & 16:11:00 +55.00.00 \\
   \textbf{Integration time} & 8~hours & 8~hours & 8~hours & 8~hours \\
   \textbf{Frequency range} & 120-166~MHz & 120-166~MHz & 115-164~MHz & 115-164~MHz \\
   \textbf{Stations (International)} & 51 (13) & 51 (13) & 50 (12) & 52 (14) \\
   \bottomrule
\end{tabular}
\label{table:data}
\tablefoot{The pointing centres were aligned during the data reduction process.}
\end{table*}


To demonstrate the SVA algorithm, we utilize calibrated datasets corresponding to four observations taken by LOFAR of the ELAIS-N1 deep field. These datasets were processed and imaged at sub-arcsecond resolution by \cite{dejong2024}. The corresponding observations are part of two different observing projects (\texttt{LT10\_012} and \texttt{LT14\_003}, PI: P.N. Best) and were downloaded from the LTA.\footnote{\url{lta.lofar.eu}} A brief summary of the metadata for these observations is provided in Table \ref{table:data}. The maximum extent between the observation times is 2.5 years.

SVA should be applied after all calibration have been performed per observation. Otherwise, various systematic or ionospheric effects cannot be properly corrected, as the original time axis information for each observation is lost and only a sidereal time axis remains.
The complete calibration process for the datasets used in this work is detailed in \cite{dejong2024}. The most essential step in the data reduction process that we need to highlight in this work, is the use of the facet-based approach to correct for DDEs \citep[e.g.][]{vanweeren2016}. Initially, direction-independent effects for the longest baselines are calibrated using one bright calibrator. Then, several other bright calibrators with sufficient S/N at the longest baselines are selected and calibrated. These calibrators define a Voronoi tessellation, where within each facet the calibration solutions are assumed to be constant \citep{schwab1984, vanweeren2016}. The visibilities from the datasets corresponding to these facets are individually imaged by subtracting sources outside these facets and then phase shifting to their centre. This process allows for additional averaging of the calibrated visibilities for each facet without introducing smearing effects, thereby speeding up the imaging process. 

In the following sections, we use datasets with calibrated visibilities from two facets to test the SVA algorithm: facet 12 and facet 25 from Figure 14 in \cite{dejong2024}. 
These datasets are stored using Dysco compression \citep{offringa2016}, which reduces data volume but is incompatible with BDA. However, as the SVA algorithm operates on a per-baseline basis, the analysis presented in the following sections is applicable to data with BDA applied as well.
Both facets were, during pre-processing, averaged to a time resolution of 8~seconds and frequency resolution of 97.66~kHz, which reduces the data volume while avoiding time and/or bandwidth smearing. Further averaging would cause smearing and lead to an irreversible loss of information that cannot be restored during SVA. The sky areas covered are 0.20~deg\(^2\) for facet 12 and 0.22~deg\(^2\) for facet 25. These correspond to the time and frequency resolution required for creating images at a resolution of 0.6\arcsec. We conduct the analysis in this paper at a resolution of  0.6\arcsec, as this offers four times better imaging speed, thereby reducing the computational resources needed to generate images. All discussions in this paper are directly applicable to a resolution of 0.3\arcsec~with the same calibrated data.

All images made in this paper are produced with \texttt{wgridder} \citep{arras2021, ye2022} from \texttt{WSClean} \citep{wsclean}, using the same settings as \cite{dejong2024}. This includes a Briggs weighting of $-1.5$ \citep{briggs1995}, a minimum \textit{uv}-value of 80$\lambda$ (corresponding to a largest angular scale of \textasciitilde43\arcmin), and a pixel size of 0.2\arcsec. For efficient deep cleaning and to better recover extended diffuse emission, we apply `auto' masking, multi-scale deconvolution, and an RMS box equal to 50 times the synthesized beam size \citep{cornwell2008, offringa2017}. Afterwards, we restore all images to a common resolution of 0.6\arcsec~to allow for direct comparisons.

\section{Results}\label{sec:results}

A key functionality of the SVA algorithm is to accurately average visibilities and weights from various observations for similar baseline coordinates. The binning is directly controlled by the given time resolution in the output dataset, as demonstrated in Figure \ref{fig:baseline_stack}. Frequency offsets between observations are not a concern, as the datasets have aligned frequency channels (see Section \ref{sec:data}), which is standard for LOFAR data (see Section \ref{sec:freqsid_axis}). The time resolution of the output dataset after applying SVA affects both the image quality, in terms of RMS noise and smearing, and the data volume. It is essential to find an optimized balance between both image quality and data volume for reducing computational time and data volume, while obtaining high-quality science-ready images after imaging. In this section, we evaluate the output of the SVA algorithm using data from the two selected facets, considering both image quality and data volume.

\subsection{Image quality}\label{sec:imagequality_res}

\begin{figure*}[htbp]
 \centering
\includegraphics[width=0.33\linewidth]{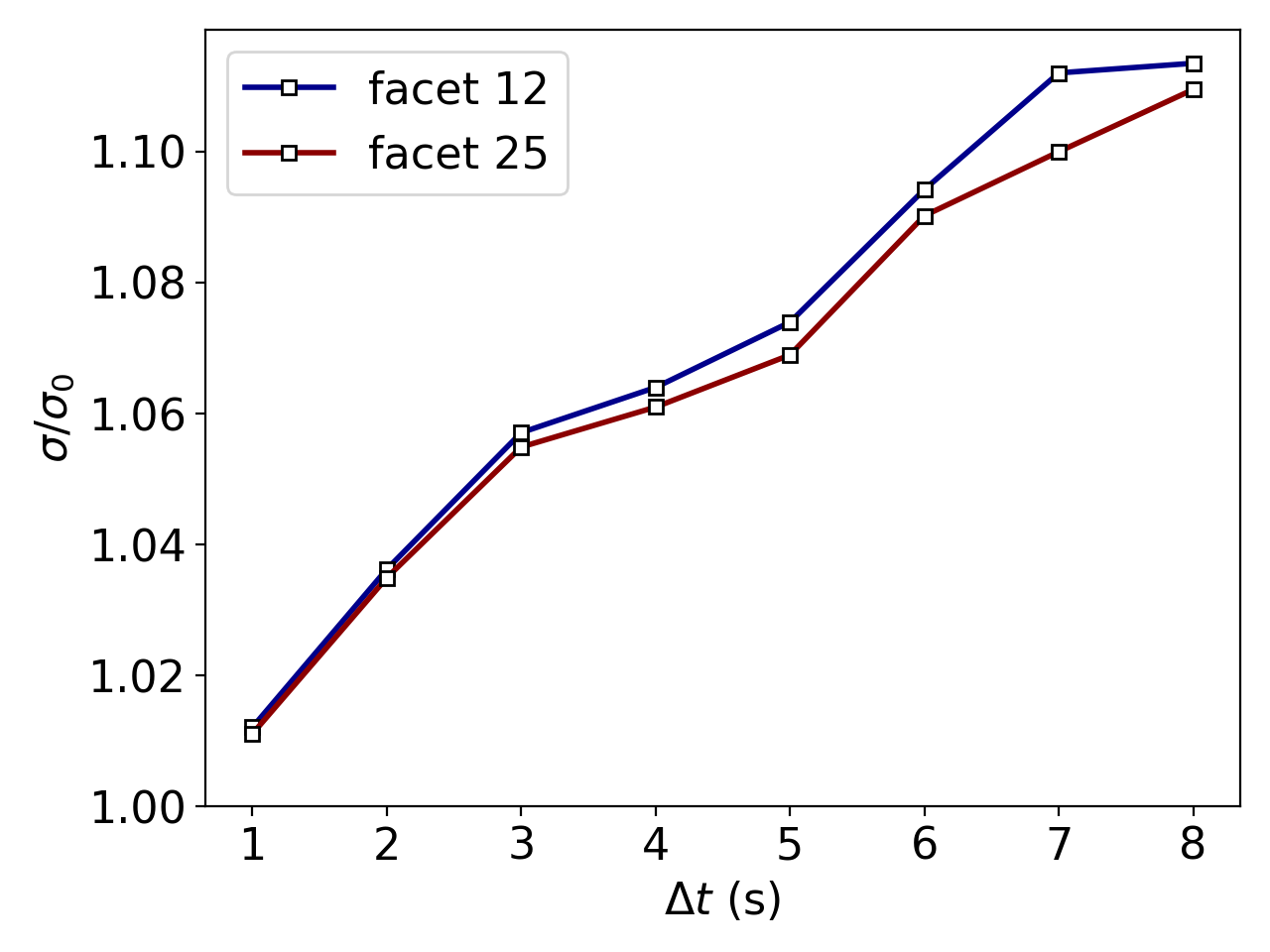}
\includegraphics[width=0.33\linewidth]{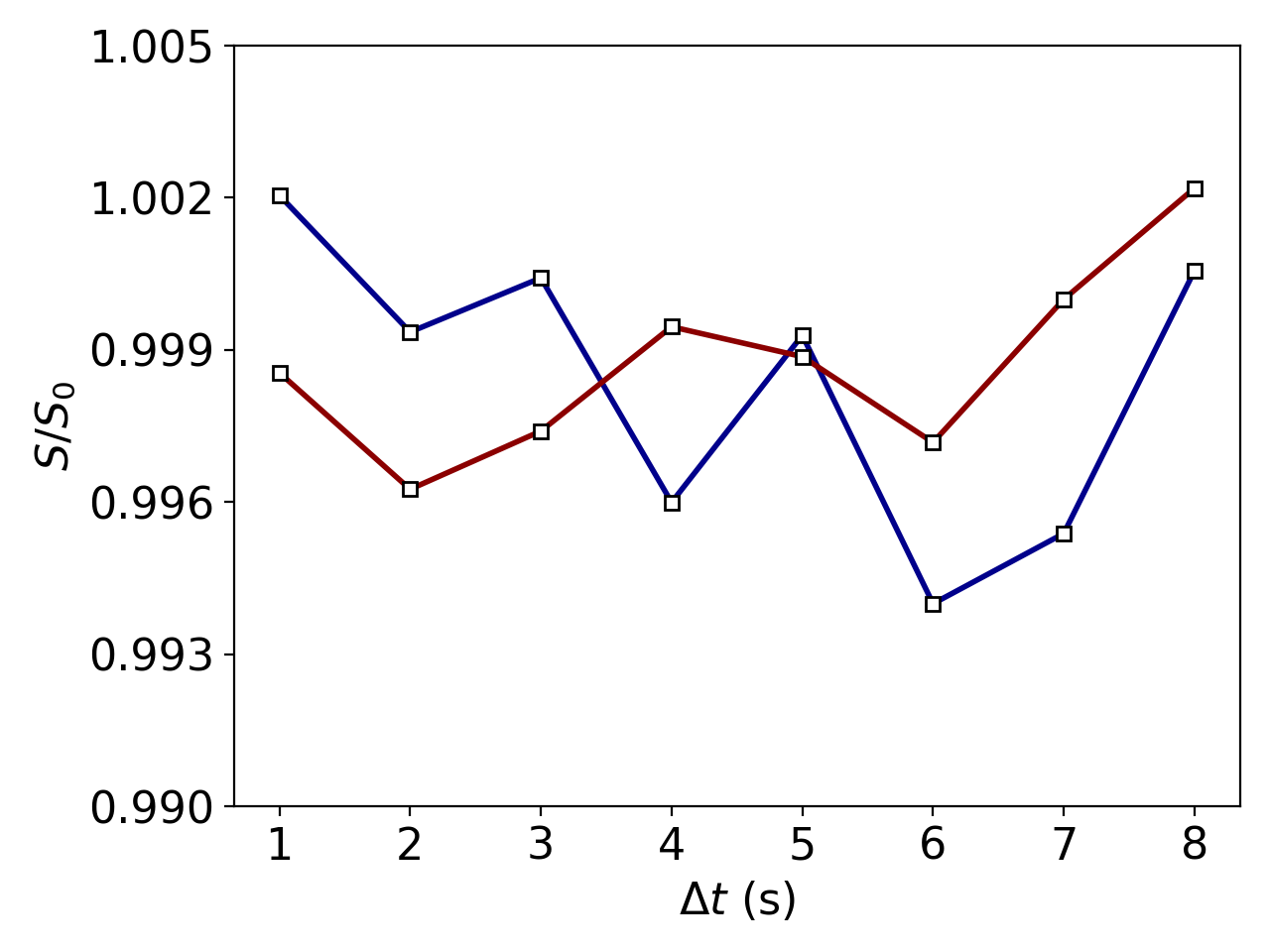}
\includegraphics[width=0.33\linewidth]{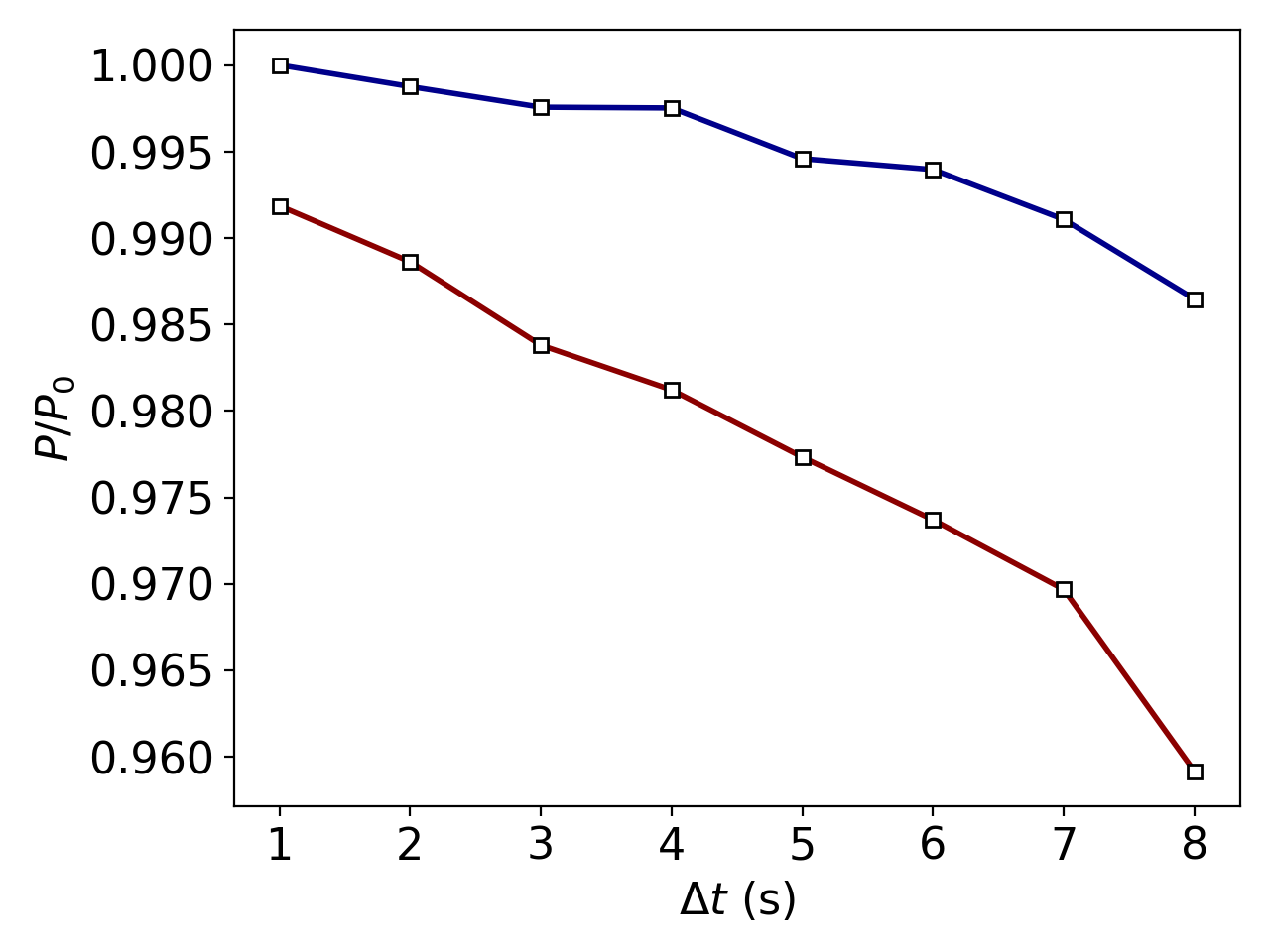}
  \caption{Comparing different image properties for different time resolutions ($\Delta t$) with and without applying SVA on datasets corresponding to facets 12 and 25 from \cite{dejong2024}. We only consider using narrower time resolutions than 8~seconds, as this was the already determined optimal time resolution to minimize data volume and prevent time smearing in the datasets prior to SVA. Any additional time averaging would lead to time smearing effects, irrespective of applying SVA. \textit{Left panel:} The measured RMS noise background ratios for each facet without ($\sigma_{0}$) and with ($\sigma$) SVA. \textit{Centre panel:} The measured integrated flux density ratios between the calibrator sources for each facet without ($S_{0}$) and with ($S$) SVA. \textit{Right panel:} The measured peak intensity ratios between the calibrator sources for each facet without ($P_{0}$) and with ($P$) SVA.}
\label{fig:time_res_background}
\end{figure*}

The datasets from the facets used in this work have both an initial time resolution of $\Delta t_{0}=8$~seconds, which was identified as a balance between data volume and time smearing by \cite{dejong2024}, when performing standard imaging without SVA. To compare the image qualities for different time resolutions of the output datasets after applying SVA, we plot in Figure \ref{fig:time_res_background} the comparison between different image properties of the original non-averaged image and the images after SVA for different time resolutions.

We find in the left panel of Figure \ref{fig:time_res_background} that the background RMS noise improves for smaller $\Delta t$. 
Facet 25 exhibits marginally lower relative RMS noise offsets because, following its DD calibration, the global amplitudes are for this facet slightly elevated compared to those in facet 12, as shown in Figure 25 of \cite{dejong2024}. 
In the centre panel of Figure \ref{fig:time_res_background}, we find no trend for the different time resolutions and offsets between the original integrated flux density ($S_0$) and the integrated flux density after SVA of less than 1\%. The right panel displays the peak intensity ratios for the calibrator sources of each facet, where we find the peak intensity to reduce up to about 5\% towards lower time resolutions.
This indicates an increase in time smearing, as smearing impacts only the peak intensity while conserving the flux densities to the first order.
This effect is more pronounced for the calibrator source of facet 25, as this source is located closer to the edge of its facet, where smearing effects are strongest.

\begin{figure*}[htbp!]
 \centering
\includegraphics[width=0.85\linewidth]{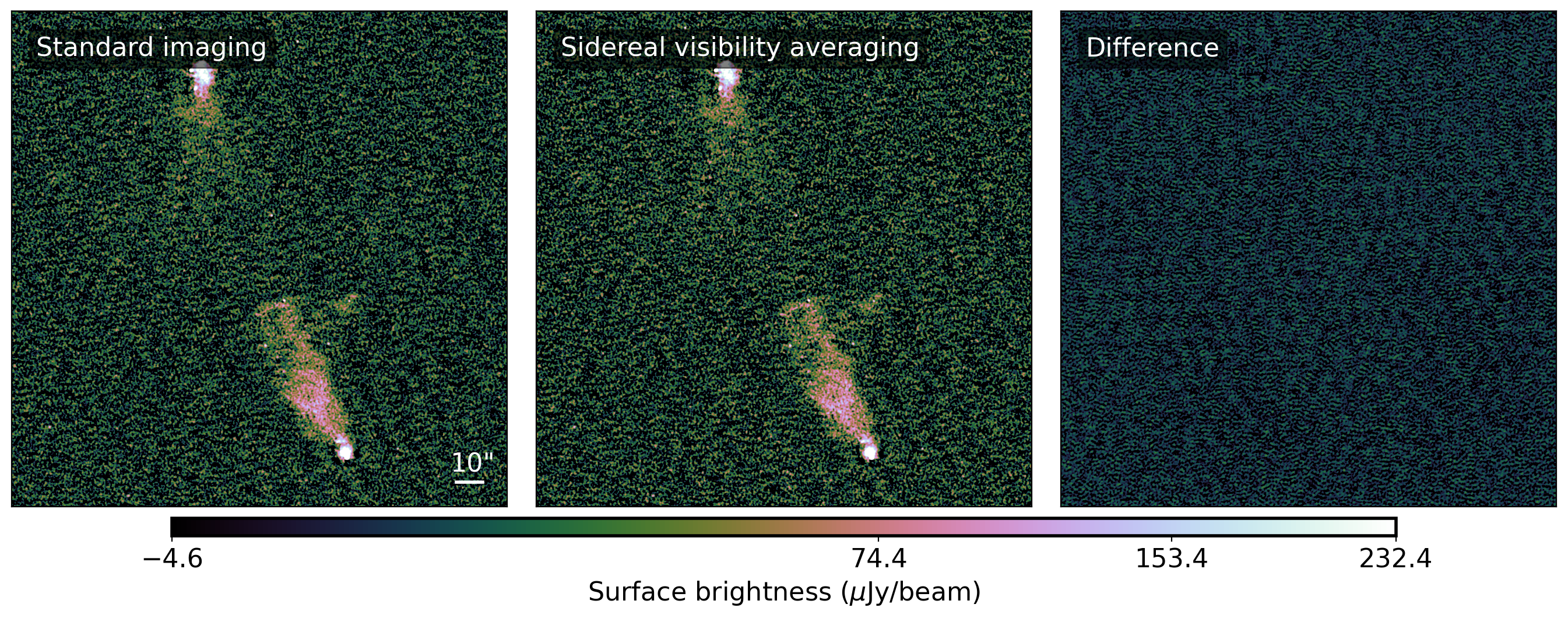}
\includegraphics[width=0.85\linewidth]{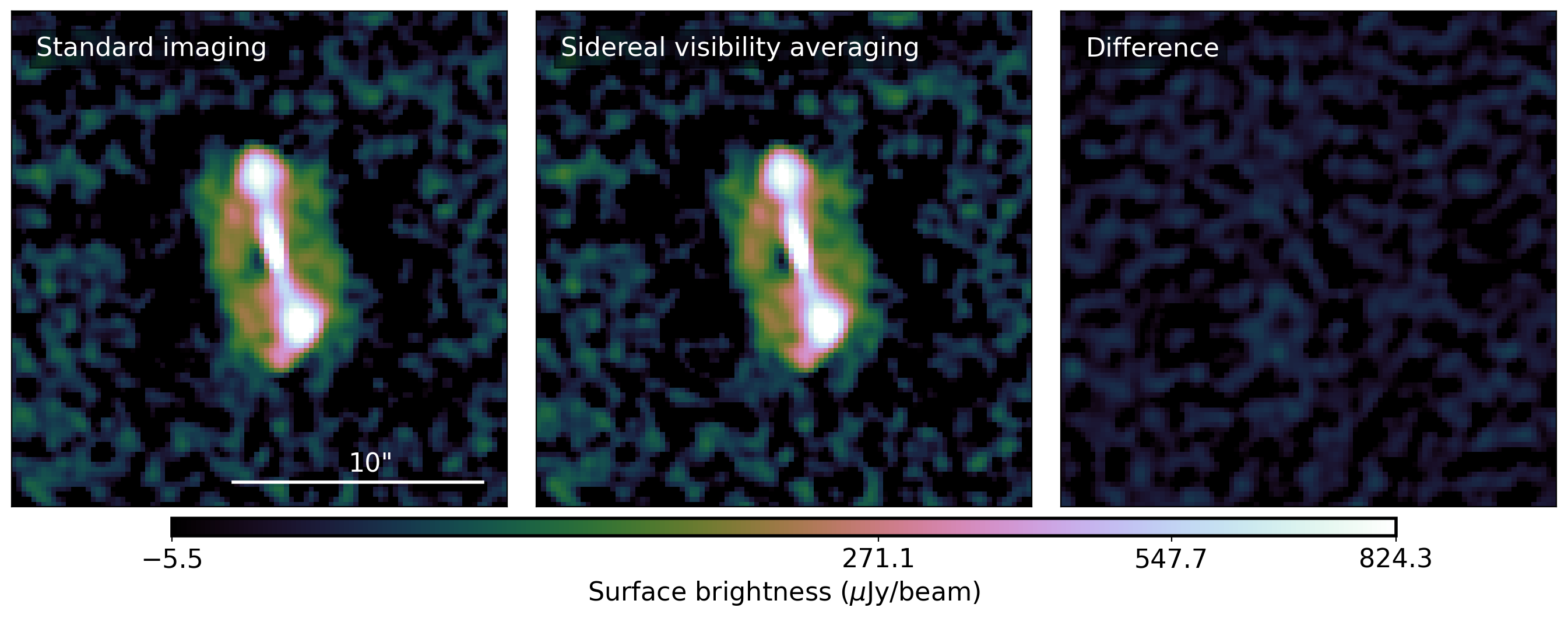}
\includegraphics[width=0.85\linewidth]{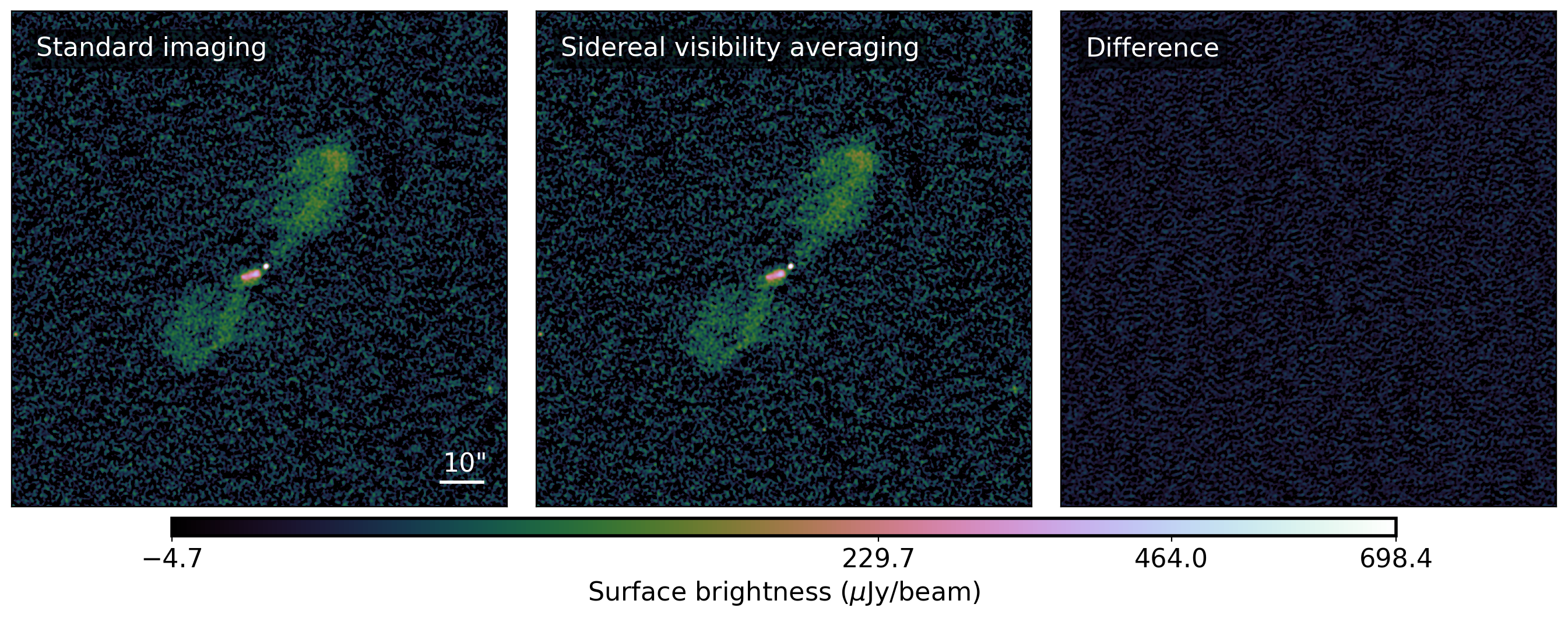}
  \caption{Comparison between the 0.6\arcsec~resolution images produced with and without SVA for three extended radio galaxies (\textit{rows}). \textit{First column:} Original image from \cite{dejong2024} produced after standard imaging without SVA. \textit{Second column:} Result when imaging after applying SVA at 4~second resolution. \textit{Third column:} The image after subtracting the second from the first image. Details about the data and imaging settings are provided in Section \ref{sec:data}.}
\label{fig:diffuse_sources}
\end{figure*}

In Figure \ref{fig:diffuse_sources}, we compare cutouts of three extended sources located in the facets when performing standard imaging, with imaging after SVA at a $\Delta t=3.6$~seconds (which follows from Equation \ref{eq:bridle} at $\tau=0.95$), and the subtraction between both images. We observe a slight increase in the RMS background noise (as shown in the left panel of Figure \ref{fig:time_res_background}), and we find as expected the sources to disappear in the subtracted images.

\subsection{Computing resources}\label{sec:computing}

The main goal of applying SVA is to reduce the required data volume and computational cost for imaging. We will therefore consider the benefits in terms of data storage and computational time.

The data volume after applying SVA ($\mathcal{V}$) depends on the sampling rate of the new time axis. If the time resolution of the new dataset matches that of the $N$ input datasets, the resulting data volume is $\mathcal{V}=\frac{\sum_{i=1}^{N} V_{i}}{N}$, where $V_{i}$ are the data volumes of the input datasets. In the more general case, where the new time resolution ($\Delta t$) differs from the input time resolution ($\Delta t_{0}$), the input and output data volumes are related by
\begin{equation}\label{eq:datvol}
\mathcal{V} \sim \frac{\Delta t_{0}}{N \cdot \Delta t} \sum_{i=1}^{N} V_{i}.
\end{equation}
This implies that to achieve any data volume reduction, it is essential that $\Delta t > \frac{\Delta t_{0}}{N}$, which gives with Equation \ref{eq:bridle}
\begin{equation}\label{eq:N_threshold}
N > 3.4\times 10^{-5}\left(\frac{\Delta t_{0}\cdot\theta}{\theta_{\text{res}}\sqrt{1-\tau}}\right),
\end{equation}
an expression for the lower bound for the number of observations that are necessary to obtain a data volume reduction, given the characteristics of the desired image size, smearing allowance, and output time resolution.
In our test case with $N=4$ observations and an input time resolution of $\Delta t_{0} = 8$~seconds, we find the data volume
\begin{equation*}
\mathcal{V} \sim \frac{2}{\Delta t} \sum_{i=1}^{4} V_{i}.
\end{equation*}
This shows that to obtain a data volume reduction (less than $\sum V_{i}$), $\Delta t > 2$~seconds is required. This is satisfied for $\tau=0.95$, as this corresponds to $\Delta t=3.6$~seconds (following from Equation \ref{eq:bridle}).

In terms of the computational costs, we know that the computational time required for imaging large datasets scales almost linearly with the number of visibilities \citep[e.g.][]{dejong2024}. This is because the amount spent on deconvolution is almost negligible with respect to the time spent on repeatedly gridding and degridding visibilities, and reading/writing them to disk. Since performing the SVA algorithm also requires additional computational time for applying the algorithm for each dataset ($T_{\text{sva}}$), we need to include this in the total computational time. This results in the following total computational time for imaging and applying the SVA algorithm:
\[
\mathcal{T} \sim T_{0} \frac{\Delta t_{0}}{N \cdot \Delta t} + N\cdot T_{\text{sva}},
\]
where $\mathcal{T}$ is the total computational time with SVA and $T_{0}$ is the total computational time when performing standard imaging without SVA.
For the SVA algorithm to be efficient, it must satisfy the condition
\[
T_{\text{sva}} < \frac{T_{0}}{N} \left(1 - \frac{\Delta t_{0}}{N \cdot \Delta t}\right),
\]
which again strictly requires $\Delta t > \frac{\Delta t_{0}}{N}$. Both our facets were originally imaged within about 42~hours without SVA. So, in this case, using 4 datasets, we find
\[
T_{\text{sva}} < 10.5 \left(\frac{\Delta t-2}{\Delta t}\right)\text{ hours}.
\] 
The SVA algorithm currently takes \(T_{\text{sva}} = 0.7\)~hour for an output time resolution of \(\Delta t = 3.6\) seconds with our 4 datasets on an Intel\textsuperscript{\textregistered} Xeon\textsuperscript{\textregistered} Gold 5220R Processor with 96 cores. This clearly satisfies the above condition and reduces the computational time by a factor of 1.6. However, it is important to emphasize that for different datasets, more optimized code, or a different processor node, the computing cost in terms of CPU~hours could be more favourable.

\section{Discussion}\label{sec:discussion}

We have introduced the SVA method and applied it on deep-calibrated LOFAR data at 0.6\arcsec~resolution. Having examined image qualities at various time resolutions, we delve in this section deeper into the image quality and computing resource balance. This enables us to estimate the advantages of scaling up to combine a larger number of observations. We also investigate the effects of Earth's celestial motion, introducing \textit{uvw} and frequency offsets.

\subsection{Image quality vs. computational resources}\label{sec:imagequality_dis}

In Section \ref{sec:imagequality_res} and Figure \ref{fig:time_res_background}, we found that image quality reduced due to increased RMS noise and increased smearing of the order of a few per cent after interpolating the input baseline coordinates to a new \textit{uvw}-plane. The trend in the left panel of Figure \ref{fig:time_res_background} also clearly shows that the RMS noise goes up for larger time resolutions.
Although Briggs weighting may in our tests have an effect on the noise difference, since the \textit{uvw} coordinates are different between the imaging with and without SVA, we still observed an increase in the RMS noise in images produced with data from a few international baselines and with uniform weighting when SVA was applied. This suggests that the primary factor behind the noise increase is related to the interpolation process, followed by the imaging where the \textit{uvw} coordinates are interpolated again onto a regular grid to facilitate efficient fast Fourier transforms (FFTs). Consequently, the gridded visibilities represent interpolations of already interpolated \textit{uvw} coordinates, compounding inaccuracies introduced during gridding.
The small loss in peak intensity when comparing data imaged with and without SVA, is attributed to additional time smearing effects. This is likely caused by the shifts in baseline coordinates introduced during SVA. This adds to smearing that may already be present in the data. 

We found that setting the time resolution to $\Delta t=3.6$ seconds, allowing a smearing factor of $\tau=0.95$, results in a computing time improvement of 1.6 times and a data volume reduction of 1.8 times when combining 4 observations of ELAIS-N1 with the current SVA code. While this is a significant gain, these resource savings increase further when combining more datasets with SVA. For example, imaging the 64 available observations of the ELAIS-N1 deep field in the LTA without SVA would be rather costly, as this would currently take about \textasciitilde1,800,000~CPU~hours for the final imaging of the data. However, using SVA this wall-time reduces by about a factor of 10, saving almost 1,600,000~CPU~hours. 
Looking even further ahead, with plans to observe a single LOFAR pointing for 3,000~hours, we anticipate reductions up to a factor of \textasciitilde169 in data volume and a 14-fold decrease in computing time, while achieving point source sensitivities in the $\mu$Jy~beam$^{-1}$ range.

The image quality reduction in terms of time smearing and RMS noise of around a few per cent compared to standard imaging, becomes in general certainly acceptable given the improved depth of the output image by $\sqrt{N}$ with approximately $\frac{\Delta t_{0}}{\Delta t}N$ lower data volumes and faster imaging times. For very large numbers of observations, one could also consider to optimize image quality by selecting a finer time resolution and more conservative smearing factors ($\tau$). In the case of 500~or 3000~hours observational time, one would for instance with $\tau=0.99$ and our current code still achieve substantial reductions in computational time by factors of approximately 7 and 12, respectively, while reducing data volumes by about 13-fold and 75-fold, respectively. Alternatively, if it is known in advance that SVA will be used to combine observations for deep imaging, these observations could be strategically scheduled to align the start and end time in LST and record the observations with the same time resolutions in LST. This ensures minimal baseline coordinate offsets between observations, disregarding, for now, the effects of Earth's celestial motion discussed in Section \ref{sec:precession}.

\subsection{Precession, nutation, and aberration}\label{sec:precession}

One of the main challenges of combining observations taken over different sidereal days, is the fact that the coordinates of baselines are not fixed and alter over time due to Earth's celestial motion. Precession is the conical motion of the Earth's rotation axis, while nutation refers to the smaller oscillations superimposed on the longer-term precession motion \citep[e.g.][]{rekier2022}. Both are for the most part due to the gravitational forces exerted by the Sun and the Moon. Precession has the most dominant effect on the baseline tracks with a rate of about 50.2\arcsec~per year, whereas nutation has a smaller effect with an amplitude of 9.2\arcsec~over a period of 18.6 years \citep{mathews2002, dehant2017}. In addition, annual aberration, which results from the Earth's orbital motion around the Sun, causes another apparent shift in the observed positions of astronomical objects. This effect introduces a maximum shift in baseline coordinates of approximately $20.5\arcsec$~over the course of a year \citep[e.g.][]{gubanov1973, kovalevsky2003}.

In Figure \ref{fig:baseline_stack}, we observe the above mentioned effects on the baseline coordinates using different time resolutions. The distances between baseline coordinates within the same dataset represent the maximum allowable separation between \textit{uvw} points, which, as determined by \cite{dejong2024}, stay within acceptable smearing limits. The different \textit{uvw} samplings, represented by the blue dots in Figure \ref{fig:baseline_stack}, demonstrate that we remain within these limits, indicating that both precession and aberration have minimal influence on the resulting dataset after SVA. However, when combining observations taken with many years in between or when for instance observing objects at higher declinations, precession becomes increasingly significant and requires adjustments to the \textit{uvw} sampling to avoid substantial time-smearing effects.

The SVA algorithm addresses the challenges introduced by celestial motions in part by using nearest-neighbour interpolation of the baseline coordinates. This can be further refined by employing a finer time resolution, as shown in the three panels of Figure \ref{fig:baseline_stack}. In some cases, it may also be more accurate to generate the output \textit{uvw}-plane through interpolation across the entire \textit{uvw} space, rather than on a per-baseline basis, since \textit{uvw} points from different baselines may overlap due to precession. When combining a large number of observations with large time intervals between them, an alternative approach is to group datasets within specified observing time ranges and apply SVA only on these subsets. This approach still reduces data volume without combining all datasets into a single set, thereby minimizing the loss of image quality caused by combining too distant \textit{uvw} coordinates from different observations. The new set of sidereal averaged datasets can then be imaged together. Additionally, to minimize aberration effects, it is advisable to schedule observations for SVA close to each other.

\subsection{Doppler shifts}\label{sec:doppler}

When we combine observations taken at different moments in time, it is also important to consider Doppler shifts which occur due to the relative motion between our instrument on earth and the sky direction, causing observed frequency changes. Doppler shift differences between observations result in frequency offsets that may spectrally distort our images. This is in particular relevant to spectral line science.

The radial velocity is given by
\[
v_r = \mathbf{v}_{\text{e}} \cdot \hat{\mathbf{r}},
\]
where \( \mathbf{v}_{\text{e}} \) is the velocity vector of the Earth and \( \hat{\mathbf{r}} \) is the unit vector pointing from the observer to the sky direction, using the pointing centre of the observation and the antenna locations relative to the centre of the Earth. The Doppler shift for an observation is
\[
\Delta \nu = \nu_{\text{obs}} \left( \frac{v_{r}}{c} \right),
\]
where $\nu_{\text{obs}}$ is the observing frequency at 140~MHz. We find Doppler shifts ranging from 2.6~kHz for the ELAIS-N1 observation with ID \textit{L686962} to -2.3~kHz for the observation with ID \textit{L769393} (see Table \ref{table:data}). Given the frequency resolutions of 97.66~kHz, these Doppler shifts are too small to significantly contribute to frequency offsets during visibility averaging. This is because ELAIS-N1 is relatively favourably positioned on the sky. When observing at lower ecliptic latitudes, Doppler shifts could reach values of up to about $\pm15$~kHz for LOFAR observations. This might have severe effects when the observations are taken half a year apart, introducing \textasciitilde30~kHz frequency offsets. 

To avoid the need to apply Doppler shift corrections when applying SVA, it may be beneficial to create for instance smaller facets. These smaller facets can be averaged more in frequency before SVA, making the Doppler shifts relatively smaller compared to the frequency channel width. Alternatively, if it is known in advance that SVA is used for imaging deep surveys, it is best to schedule the observations strategically. This way, observations can be scheduled to minimize the introduction of large Doppler shifts between the individual observations. A potentially better solution is to apply default Doppler corrections during the measurement pre-processing phase, before any visibility data is calibrated. This approach would eliminate the need to account for Doppler corrections during data processing. However, this may also need to involve additional corrections for \textit{uvw} coordinates.

\section{Summary and conclusion}\label{sec:conclusion}

We have in this paper revisited a method called `sidereal visibility averaging' (SVA) to enable ultra-deep imaging when combining multiple observations of a single pointing on the sky. This method takes advantage of the repetitive nature of baselines each sidereal day, allowing us to average calibrated visibilities from different observations at similar baseline coordinates. While this approach eliminates information about the time-varying flux of transient sources, it significantly reduces the number of visibilities to process during imaging, alleviating the computational bottleneck for deep imaging with multiple observations and lowering the long-term data archiving costs of calibrated visibilities. It can be used in addition to other data volume reduction methods, such as BDA and Dysco compression.

By testing the SVA algorithm with four previously calibrated datasets from \cite{dejong2024}, corresponding to images of two facets at a 0.6\arcsec~resolution, we found that we could reduce the data volume by a factor of 1.8 and speed up imaging by a factor of 1.6 compared to standard imaging when we allow a 5\% additional smearing increase towards the edge of the imaged facet. The improvements in data volume and computational time become larger when more observations are combined. For example, applying this method to the approximately 500~hours of LOFAR data available for the ELAIS-N1 deep field, we estimate reductions in data volume of up to a factor of 28, while computing times may be decreased by a factor of around 10. For even larger projects, with over 3000~hours of combined integration time, the improvements may reach up to a factor of 169 in data volume and a factor of 14 in computing time, while achieving imaging sensitivity of the order of a $\mu$Jy~beam$^{-1}$ at 150 MHz. The computational time reductions are likely to further improve as the software becomes more optimized and more advanced hardware becomes available.

We also examined the effects of Earth's celestial motion on the baseline coordinates of the combined dataset. Although the baseline coordinates between the four observations are offset due to precession and aberration, these have a small effect on SVA. However, we anticipate that these effects could become problematic when combining observations taken many more years apart. Depending on the resolution, this could potentially introduce additional smearing effects. This issue can be addressed by adding more \textit{uvw} points to the output dataset using finer time resolutions, which reduces the final computational speedup factors but maintain image quality. We also assessed the effects of Doppler shifts and found that for the utilized ELAIS-N1 data, they are too small to have a significant impact on the dataset after applying SVA. However, observations at different positions on the sky may necessitate additional frequency corrections, depending on the times of the year during which the observations were conducted. Ideally, to optimize baseline coordinate binning accuracy and minimize Doppler shift effects, observations should be scheduled during the same period of the year and start at the same sidereal time.

We have demonstrated that SVA is an effective method for reducing data volumes and processing time for imaging calibrated visibilities. We believe this approach serves as an important building block for producing the deepest single-pointing images using current and upcoming interferometric radio instruments, as long the calibrated visibility data of the observations being combined are retained before applying SVA. This enables us to explore the universe at unprecedented depths and spatial resolution.

\begin{acknowledgements} 

This publication is part of the project CORTEX (NWA.1160.18.316) of the research programme NWA-ORC which is (partly) financed by the Dutch Research Council (NWO). This work made use of the Dutch national e-infrastructure with the support of the SURF Cooperative using grant no. EINF-1287. This work is co-funded by the EGI-ACE project (Horizon 2020) under Grant number 101017567.
 
RJvW acknowledges support from the ERC Starting Grant ClusterWeb 804208. 

FS appreciates the support of STFC [ST/Y004159/1].

LOFAR data products were provided by the LOFAR Surveys Key Science project (LSKSP; \url{https://lofar-surveys.org/}) and were derived from observations with the International LOFAR Telescope (ILT). LOFAR \citep{haarlem2013} is the Low Frequency Array designed and constructed by ASTRON. It has observing, data processing, and data storage facilities in several countries, which are owned by various parties (each with their own funding sources), and which are collectively operated by the ILT foundation under a joint scientific policy. The efforts of the LSKSP have benefited from funding from the European Research Council, NOVA, NWO, CNRS-INSU, the SURF Co-operative, the UK Science and Technology Funding Council and the Jülich Supercomputing Centre.

\end{acknowledgements} 

\bibliographystyle{bst}
\bibliography{bib}

\appendix

\end{document}